\documentstyle[12pt]{article}


\def\Lp{\displaystyle{\biggl(}}
\def\Rp{\displaystyle{\biggr)}}

\newcommand{\G}{\Gamma}
\newcommand{\D}{\Delta}

\renewcommand{\b}{\beta}
\renewcommand{\d}{\delta}
\newcommand{\e}{\varepsilon}

\newcommand{\g}{\gamma}

\renewcommand{\l}{\lambda} 
\newcommand{\m}{\mu}
\newcommand{\n}{\nu}

 \renewcommand{\O}{\Omega}

\newcommand{\s}{\sigma} \renewcommand{\S}{\Sigma}

\newcommand{\DD}{{\cal D}}

\newcommand{\GG}{{\cal G}}

\newcommand{\SS}{{\cal S}}

\newcommand{\UU}{{\cal U}}

\newcommand{\WW}{{\cal W}}

\newcommand{\complex}{{\kern .1em {\raise .47ex
\hbox {$\scriptscriptstyle |$}}
    \kern -.4em {\rm C}}}
\newcommand{\real}{{{\rm I} \kern -.19em {\rm R}}}
\newcommand{\rational}{{\kern .1em {\raise .47ex
\hbox{$\scripscriptstyle |$}}
    \kern -.35em {\rm Q}}}
\renewcommand{\natural}{{\vrule height 1.6ex width
.05em depth 0ex \kern -.35em {\rm N}}}
\newcommand{\dint}{\displaystyle{\int}}

\newcommand{\cb}{{\bar c}}

\newcommand{\pa}{\partial}

\newcommand{\dfrac}[2]{{\displaystyle{\frac{#1}{#2}}}}

\newcommand{\sla}{\raise.15ex\hbox{$/$}\kern -.57em}

\newcommand{\twiddle}{\lower.9ex\rlap{$\kern -.1em\scriptstyle\sim$}}

\newcommand{\equ}[1]{(\ref{#1})}

\newcommand{\eq}{\begin{equation}}
\newcommand{\eqn}[1]{\label{#1}\end{equation}}
\newcommand{\eea}{\end{eqnarray}}
\newcommand{\eqa}{\begin{eqnarray}}
\newcommand{\eqan}[1]{\label{#1}\end{eqnarray}}
\newcommand{\ba}{\begin{array}}
\newcommand{\ea}{\end{array}}
\newcommand{\eqac}{\begin{equation}\begin{array}{rcl}}
\newcommand{\eqacn}[1]{\end{array}\label{#1}\end{equation}}

\newcommand{\at}{{\~a}}

\newcommand{\ooo}{{\'o}}

\newcommand{\iii}{{\'\i}}

\begin{document}


{\ }

\vspace{3cm}
\centerline{\LARGE Duality Mechanism by Introduction of }\vspace{2mm}
\centerline{\LARGE Gauge Condensates }

\vspace{1cm}
\centerline{\bf {\large M.A.L.Capri, V.E.R.Lemes}}
\vspace{2mm}
\centerline{{\it UERJ - Universidade do Estado do Rio de Janeiro}}
\centerline{{\it Instituto de F{\iii}sica }}
\centerline{{\it Departamento de F{\iii}sica Te{\ooo}rica}}
\centerline{{\it Rua S{\at}o Francisco Xavier, 524}}
\centerline{{\it 20550-013 Maracan{\at}, Rio de Janeiro, Brazil}}
\vspace{4mm}

\centerline{{\normalsize {\bf REF. UERJ/DFT-09/2004}} }

\vspace{4mm}
\vspace{10mm}

\centerline{\Large{\bf Abstract}}\vspace{2mm}
\noindent In this work, we study the implications of the existence of a gauge condensate to the mechanism of duality, a method based on the existence of these condensates is presented and applied to the study of the dual equivalence between self-dual (SD) and topologically massive Yang-Mills (TMYM) models.

\newpage
\section{Introduction}

\noindent\hspace{0.75cm}It has been pointed out by many authors the duality between the dimensional Maxwell-Chern-Simons
and Self-Dual Abelian \cite{Deser,Nieu} and the duality between the Yang-Mills-Chern-Simons and non-Abelian Self-Dual \cite{Bralic,Town} models. The central aspect of this duality is the number of degrees of freedom and consistent ways to implement them \cite{Ribeiro,Ilha,gomes,clovis}. In this work, we deal with the same issue as these works but concentrate in the introduction of mass for the gauge fields,which allows the possibility of generalization of the duality mechanism to another dimensions \cite{Botta}.

\noindent\hspace{0.75cm}It is well know that topological actions are a class of gauge models with the property that their observables are of topological nature and there is no intrinsic scale or metric observables in such models. On the other hand, the duality equivalence, which was best exemplified by the one between the self-dual (SD) and the Yang-Mills Chern-Simons (YMCS) model \cite{Ribeiro,Ilha,gomes,clovis,Botta} is under two massive actions. One point needs more attention: the introduction of the mass  parameter in order to obtain the (SD) model and stablish the duality equivalence with the (YMCS)\cite{Botta2}. It is usual to admit that the mass could come from the bosonisation process over certain fermionic action. Our point of view is a little bit different. We use the topological actions as starting point actions without degrees of freedom and introduce the mass term, as a part of the requirements for the insertion of a field condensate \cite{K.Knecht,Van}, in order to obtain the correct number of liberty degrees. In this sense, an insertion that is quadratic in a gauge field (or fields) give us an equation of motion (for such field) that can be used to obtain the dual action. The mechanism of use the equation of motion is already well established and used into many other approaches to obtain the dual action \cite{Ribeiro,Ilha}. The main difference of our approach is the use of the gauge insertion procedure in order to introduce the mass into a consistent way, fixing the number of degrees of freedom by the introduction of the condensates, and ensuring the stable starting point action in order to use properly the equation of motion.

\noindent\hspace{0.75cm}The paper is organized as follows: In section {\it II} we describe the method beginning with a brief review about the main properties of topological actions. The gauge field insertion mechanism is presented and his relations with the degrees of freedom and the stability of the action is discussed.  We devote section {\it III} to present some examples of such method in implementing the duality between three-dimensional gauge theories in special the duality between the self-dual (SD) and the Yang-Mills Chern-Simons (YMCS) model. We also present one example in which there is an obstruction to implement duality \cite{Rodrigues}.

\section{Duality Mechanism and Gauge Condensates.}

\noindent\hspace{0.75cm}In order to present the main ideas of the method let us first remember fundamental properties of fully quantized topological type actions, in particular the Chern-Simons and BF ones. These theories provide examples of completely finite quantum theories, i.e., theories with both vanishing $\b$-function and anomalous dimension. In the functional point of view, the latter property is given by the vector supersymmetry that, with the BRST operator, generates a Wess-Zumino type algebra \cite{livro}:

\eq
\WW^{\m}\Sigma = {\Delta^{\m}}_{cl}, \hspace{1cm} \ss_\S\Sigma = 0,\hspace{1cm} \left\lbrace \ss_\S,\WW^{\m} \right\rbrace = \partial^{\m}, 
\eqn{topo}

\noindent where $\S$ is a given topological action, $\ss_\S$ is the linearized nilpotent operator that extends the BRST symmetry to all fields and sources of  the action and ${\cal W}^\m$ is a Ward operator that, together with $\ss_\S$ generates the translation Wess-Zumino algebra .  ${\Delta^{\m}}_{cl}$ is a classical breaking.

\noindent\hspace{0.75cm}These topological actions provide the main support for the introduction of gauge insertions, using BRST doublets:

\eq\ba{ll}
\widetilde\S[\varphi]=\S[\varphi]
+\ss\biggl(\dint d^d\!x\,\l f(\varphi)\biggr)=
\S[\varphi]+\dint d^dx \,\Bigl[\chi f(\varphi)-\l\ss\Bigl(f(\varphi)\Bigr)\Bigr], \\  \\
\hspace{3cm}
\ss\l=\chi,\qquad\ss\chi=0,
\ea\eqn{breaks}

\noindent Where $f(\varphi)$ corresponds to the insertions in the gauge fields $\varphi$, and $(\l,\chi)$ corresponds to a BRST doublet. This insertion explicitly breaks the vector supersymmetry and the Wess-Zumino algebra, ensuring that the number of degrees of freedom is determined by the insertion. It is important to say that the BRST symmetry is not broken, these result is fundamental for the explicitly calculation of the mass value associated to insertions in any dimensions \cite{Gracey,Dudal}. Taking into account this requirement, the introduction of the gauge insertion is done in such a way that the equation of motion over this insertion generates one linear term in the field. This type of insertion gives to the $\chi$ source canonical dimension $d-2[dim(\varphi)]$. Taking into account that $\chi$ has non zero value when $dim(\varphi)\leq\frac{d}{2}$ \cite{K.Knecht,Acoleyen}, the $\chi$ source can be splited into a source $j$, with $\langle j\rangle = 0$, and a mass

\begin{eqnarray}
\widetilde\S[\varphi]\hspace{-0.22cm}&=&\hspace{-0.22cm}\S[\varphi]+\dint d^3\!x\,\biggl[\dfrac{1}{2}(j+m)\varphi^2-\l\varphi(\ss\varphi)\biggr]\label{3dinsertion}\\
\dfrac{\d\widetilde\S}{\d\varphi}\hspace{-0.22cm}&=&\hspace{-0.22cm}\dfrac{\d\S}{\d\varphi}+(j+m)\varphi-\l(\ss\varphi)-\l\varphi\dfrac{\d}{\d\varphi}(\ss\varphi).
\label{eqmov} 
\end{eqnarray}

It is important to emphasize here the need of a functional equation that controls the insertion. Such equation fixes the gauge fixing choice that is compatible with the insertion. In $3$ dimensions this equation has the functional form
\eq
\dint d^3\!x\,\biggl[\dfrac{\delta\widetilde{\Sigma}}{\delta\lambda}] +c\dfrac{\delta\widetilde{\Sigma}}{\delta b}\biggr]=0
\eqn{eq-insert}
Where $b$ is the lagrange multiplier that defines the gauge fixing.

This functional equation with the Slavnov-Taylor identity, generates the $\delta $ symmetry. In four dimensions, the $\delta $ equation is the functional form of the SL(2R) symmetry \cite{Lemes}. Thus the $\delta $ equation together with the insertion equation are both responsible for the stability of the insertion into the starting point action. With a stable starting point action, the use of the equation of motion gives the dual action, 

\eq 
\widetilde{\Sigma}[\varphi]\longrightarrow \varphi = 
\dfrac{1}{m}(\mbox{eq. mov.})\biggl|_{(\lambda,j)=0} 
\longrightarrow \Xi [\widetilde{\varphi}].
\eqn{redef}

The requirement of stability of the starting point action ensure that no other terms are generated by quantum corrections to the equation of motion,

\eq\ba{ll}
\widetilde{\Sigma}[Z\varphi]\equiv\G[\varphi]=\widetilde{\Sigma}[\varphi] + \hbar \Gamma_{c}[\varphi]+O(\hbar^2), \vspace{0.25cm}\\
\varphi_0=Z\varphi = \varphi + \hbar z \varphi+O(\hbar^2),\vspace{0.25cm}\\
\dfrac{\delta\widetilde{\Sigma}[Z\varphi]}{\delta \varphi} = 
\dfrac{\delta\G[\varphi]}{\delta \varphi}.
\ea\eqn{estable}

\noindent Where $\G[\varphi]$ is the quantum action and $\G_c[\varphi]$ is the counter-term action.

\section{$3$ Dimensional Examples}
 
\noindent\hspace{0.75cm}For a better understanding of the method, let us work up some examples beginning by the case of the duality between the Nonlinear Self-Dual (NSD) and Topologically Massive Yang-Mills (TMYM) models. In order to point out the limits of the duality mechanism, the duality involving BF plus Chern-Simons (BF+CS) and Maxwell Chern-Simons is also presented and the impossibility of the extension to the non-abelian case is discussed. 

\subsection{Selfdual and Topologically Massive case.}

\noindent\hspace{0.75cm}The starting point of the duality mechanism applied to the NSD model is the fully quantized action $ \Sigma $ that contains the insertion of $A^{a}_{\m}A^{a\m}$. This insertion has canonical dimension $ 2 $ and generates a linear term in the equation of motion for the $A^{a}_{\m}$ field. The action $\Sigma$ is given by:

\eq\ba{ll}
\S=\dint d^{3}\!x\biggl[\dfrac{1}{4}\e^{\m\n\s}\biggl( A^{a}_{\s}F^{a}_{\m\n}
- \dfrac{1}{3}g f^{abc}A^{a}_{\m}A^{b}_{\n}A^{c}_{\s}\biggr)
+ b^{a} \pa^{\m}\!A^{a}_{\m} + \cb^{a} \pa_{\m}(D^{\m}c)^{a}+\vspace{0.25cm}\\
\hspace{2.0cm}- \O^{a}_{\mu}(D^{\mu}c)^{a}
+\dfrac{1}{2}gf^{abc}L^{a}c^{b}c^{c}
+\dfrac{1}{2}(j+m)A^{a}_{\m}A^{a\m}+\l \pa_{\mu}c^{a}A^{a\mu}+\vspace{0.25cm}\\
\hspace{2.0cm}+ \tau_{1}m^2j+\dfrac{\tau_{2}}{2} m j^{2}
+ \dfrac{\tau_{3}}{6}  j^{3}\biggr].
\ea\eqn{sigma}

\noindent The last three terms that appear in the starting point action are only to complete the most general power-counting invariant action with all sources and fields. They are only responsible for the renormalization of the coefficients $\tau_{1}$, $\tau_{2}$ and $\tau_{3}$. 

\noindent\hspace{0.75cm}The next step is to obtain all symmetries that can be extended to quantum level \cite{livro}. These symmetries are in the form of Ward identities and are given by:

\vspace{0.5cm}

\noindent The Slavnov-Taylor identity 
\eq
\SS (\S ) = \int d^{3}\!x \Lp \dfrac{\d \S}{\d\! A^{a}_{\m}}\dfrac{\d \S}{\d \O^{a \, \m}}
+ \dfrac{\d \S}{\d c^{a}}\dfrac{\d \S}{\d L^{a}} + b^{a}\dfrac{\d \S}{\d \cb^{a}}
+ (j + m)\dfrac{\d \S}{\d \l}  \Rp = 0,
\eqn{slavnov}

\noindent and the equation that defines the insertion, 

\eq
\UU(\S ) = \int d^{3}\!x \Lp \dfrac{\d \S}{\d \l} + c^{a}\dfrac{\d \S}{\d b^{a}} \Rp =0.
\eqn{insert}

\noindent This equation is strongly related to the gauge fixing choice and in our case is implemented with the Landau gauge,

\eq
\dfrac{\d \S}{\d b^{a}} = \pa^{\mu}\!A^{a}_{\m},\qquad\dfrac{\d \S}{\d \cb^{a}} + \pa^{\m}\dfrac{\d \S}{\d \O^{ a \, \m}}=0.
\eqn{gf-aght}

\noindent The insertion equation, together with the Slavnov-Taylor identity are responsible for generating the $\delta $ identity,

\eq
\DD(\S ) = \int d^{3}\!x \Lp c^{a}\dfrac{\d \S}{\d \cb^{a}} + \dfrac{\d \S}{\d L^{a}}\dfrac{\d \S}{\d b^{a}} \Rp =0.
\eqn{sl2r}

The  $\delta $ and the insertion identity are responsible for stabilizing the insertion  $A^{a}_{\m}A^{a\m}$. The Landau gauge fixing also has the ghost equation,

\eq\ba{ll}
\hspace{5cm}\GG^{a} \S = - \D^{a},\nonumber \\  \nonumber \\
\GG^{a} = \dint d^{3}x \Lp \dfrac{\d\enskip }{\d c^{a}} + gf^{abc} \cb^{b}\dfrac{\d\enskip }{\d b^{c}}  \Rp\qquad
\D^{a} = \dint d^{3}x[gf^{abc}(A^{b}_{\m}\O^{c \, \m} + L^{b}c^{c})],
\ea\eqn{def-ghost}

\noindent this equation is responsible for the nonrenormalizability of the ghost \cite{livro}. The last functional symmetry is the Rigid one,

\eq
\WW^{a}(\S)  = \int d^{3}\!x\,g f^{abc}\Lp 
A^{b}_{\m}\dfrac{\d \S}{\d\! A^{c}_{\m}} + \O^{b}_{\m}\dfrac{\d \S}{\d \O^{c}_{\m}} + c^{b}\dfrac{\d \S}{\d c^{c}} + L^{b}\dfrac{\d \S}{\d L^{c}} 
+ \cb^{b}\dfrac{\d \S}{\d \cb^{c}} 
+ b^{b}\dfrac{\d \S}{\d b^{c}}\Rp = 0. 
\eqn{Rigida}

\noindent\hspace{0.75cm}This is the complete set of equations for the study of the stability of the starting point action. Now for further use let us display the quantum numbers of the fields and sources presented in the starting point action.

$$
\begin{tabular}{|l|l|l|l|l|l|l|l|l|}
\hline
\/& $A_{\mu}^a$ &  $c^{a}$ & $\,\overline{c}^{a}$  
& $b^{a}$ & $\,\,\lambda$ & $j$ & $\Omega_{\mu}^{a}$ & $L^{a}$ \\
\hline
Gh. number & \thinspace 0 & \thinspace 1 & $-1$ & 0 & $-1$
& 0 & $-1$ & $-2$ \\
\hline
Dimension & \thinspace 1 & $\,\,$0 & $\,$1 & $\,$1 & $\,\,$1 & 1 & $\,$2 &
$\,3$ \\
\hline
\end{tabular}\\
$$
\begin{center}
{Table1: Ghost number and canonical dimension of the fields and sources.}
\end{center}

\noindent\hspace{0.75cm}The next step, consists in using this set of symmetries to obtain the most general counterterm action, $\G_c$, that can be freely added to the starting point action, $\Sigma$, in order to obtain the quantum action, $\G$, with all quantum corrections. This counterterm action is an arbitrary integrated local functional of canonical dimension $3$ and ghost number $0$. 

\noindent\hspace{0.75cm}The requirement that the perturbed action, $\S+\hbar\G_c$, obeys the same set of equation obtained for $\S $ give rise to the full set of constraints under the quantum action $\G $:

\begin{equation}
\begin{tabular}{lll}
$\SS (\G) = 0,\hspace{0.5cm}$&$\UU(\G ) =0,\hspace{0.5cm}$&$\GG^{a}(\G) = -\D^{a}\cdot\G,\vspace{0.25cm}$\\
$\DD(\G)=0,\hspace{0.5cm}$&$\dfrac{\d \G}{\d b^{a}} = \pa^{\mu}A_{\mu}^{a},\hspace{0.5cm}$&$\dfrac{\d \G}{\d\bar c^{a}} + \pa^{\m}\dfrac{\d \G}{\d \O^{ a \, \m}} = 0.\vspace{0.25cm}$
\end{tabular}
\label{vinculos}
\end{equation}

\noindent\hspace{0.75cm}The constrains over the Counterterm are obtained using the relation $\G=\S+\hbar\G_c$, the set of constraints over $\S$ and \equ{vinculos}:

\begin{equation}
\begin{tabular}{lll}
$\ss_\S(\G_c) = 0,\hspace{0.5cm}$&$\UU(\G_c) =0,\hspace{0.5cm}$&$\GG^{a}(\G_c) =0,\vspace{0.25cm}$\\
$\DD_\S(\G_c)=0,\hspace{0.5cm}$&$\dfrac{\d \G_c}{\d b^{a}}=0 ,\hspace{0.5cm}$&$\dfrac{\d \G_c}{\d\bar c^{a}} + \pa^{\m}\dfrac{\d \G_c}{\d \O^{ a \, \m}} = 0.\vspace{0.25cm}$
\end{tabular}
\label{vinculos2}
\end{equation}

\noindent Where the linearized nilpotent BRST-operator $\ss_\S$ and $\DD_\S$ are given by:

\eq
\ss_{\S} = \dint d^{3}\!x \Lp \dfrac{\d\S}{\d\!A^{a}_{\m}}\dfrac{\d\quad }{\d \O^{a\m}}
+ \dfrac{\d \S}{\d \O^{a \, \m}}\dfrac{\d\quad }{\d\! A^{a}_{\m}}
+ \dfrac{\d \S}{\d c^{a}}\dfrac{\d\quad }{\d L^{a}} + \dfrac{\d \S}{\d L^{a}}\dfrac{\d\enskip }{\d c^{a}} + b^{a}\dfrac{\d\enskip }{\d \cb^{a}}
+ (j + m)\dfrac{\d \enskip}{\d \l}\Rp,
\eqn{linearizado}

\begin{equation}
\DD_\S=\dint d^3\! x\Lp c^a\dfrac{\d\,\,}{\d\bar c^a}+\dfrac{\d\S}{\d L^a}\dfrac{\d\,\,}{\d b^a}+\dfrac{\d\S}{\d b^a}\dfrac{\d\,\,\,}{\d L^a}\Rp.
\label{sl2R}
\end{equation}

\noindent\hspace{0.75cm}After the use of all the constraints \equ{vinculos2}, we obtain the most general counterterm action:

\eq\ba{ll}
\G_c=\dint d^3\!x\biggl[\dfrac{\sigma+2a_1}{4}\e^{\m\n\s}A^a_\s F^a_{\m\n}-\dfrac{\sigma}{12}\,g\e^{\m\n\s}f^{abc}A^a_\m A^b_\n A^c_\s+a_1(\O^a_\m+\pa_\m\bar c^a)\pa^\m c^a+\vspace{0.25cm}\\
\hspace{1cm}+\dfrac{a_1}{2}(j+m)A^{a\m}A^a_\m+2a_1\l\pa_\m c^a A^{a\m}+\dfrac{a_2}{2}m^2j +\Bigl(\dfrac{a_2}{2}+\dfrac{a_3}{6}\Bigr)m j^2+\dfrac{a_3}{6}j^3\biggr].
\ea\eqn{count-act}

\noindent\hspace{0.75cm}The 4 free parameters $(\s, a_{1}, a_{2}, a_{3})$ present in the counterterm action are responsible
for the renormalization of all fields, sources, the coupling constant $g$ and parameters $\tau_{i}$.

\noindent\hspace{0.75cm}It is now clear, that the starting point action is stable and we do not have new terms generated by quantum corrections. The results obtained by the equation of motion in the starting point action $ \Sigma $ remains and the unique possible corrections comes from the multiplicative renormalization of all fields, sources, the coupling constant and parameters. This affirmative is equivalent to say that the duality obtained by direct use of the relations in the equation of motion is preserved at quantum level. Now the use of the equation of motion, for the $A^a_\m $ field, give rise to the following relation:

\eq
A^{\m \, a} = -\dfrac{1}{m}\Lp \dfrac{1}{2}\e^{\m\n\s}F^{a}_{\n\s} - \partial^{\m}b^{a} + gf^{abc}(\Omega^{\m \, b} + \partial^{\m}\overline{c}^{b} )c^{c} + jA^{\m \, a} + \lambda\partial^{\m}c^{a} - \dfrac{\delta\Sigma}{\delta\! A^{a}_{\m}}\Rp .
\eqn{eqmovCSdu}

\noindent\hspace{0.75cm} The relation, with $(j, \lambda)=0$ corresponds to the one that implement the duality. It is important to say that even setting $(j, \lambda)=0$ in the starting point action, is possible to find a functional closed form for the Slavnov-Taylor identity, using the fact that the BRST variation of the insertion 
$\dint d^3\!x\dfrac{1}{2}({A^a}_{\m}A^{a \,\m})$ is equal to $\dint d^3\!x(c^{a}\dfrac{\delta\Sigma}{\delta b^{a}})$. The Slavnov-Taylor identity without the sources $j,\lambda$ is not nilpotent and closes over the $\delta$ symmetry. In any case this identity is not broken. The other source $\Omega^{\m \, b}$ that appear into this equation of motion is necessary only to renormalise the BRST of the field ${A^a}_{\m}$ and can be set to zero at the end of all calculations.

\noindent\hspace{0.75cm}Another important example is the case of the duality between the BF+CS action and the nonabelian YMTM model. This is a good example of restrictions into implementing the duality. 

\subsection{BF plus Chern-Simons and restrictions into implement duality with YMTM model.}

\noindent\hspace{0.75cm}The Fully quantized action for the nonabelian BF+CS case with all fields and the insertions of $ {B^{a}}_{\m}B^{a\,\m} $ and $ {A^{a}}_{\m}A^{a\,\m} $ is given by:

\begin{equation}\begin{array}{ll}
S=\dint d^3x\biggl\{ \dfrac{1}{2}B^a_{\m}\e^{\m\n\s}F^{a}_{\n\s} 
+ \dfrac{1}{4}\e^{\m\n\s} \biggl( A^{a}_{\s}F^{a}_{\m\n}
- \dfrac{1}{3}g f^{abc}A^{a}_{\m}A^{b}_{\n}A^{c}_{\s} \biggr) \vspace{0.25cm}\\ \hspace{1.75cm} + b^a\partial^\m A^a_{\m} + \overline{c}^{a}\partial_{\m}D^{\m}c^{a} + h^aD^{\m}B^a_{\m} + \overline{\xi}^{a}D^{\m}D_{\m}\xi^{a} \vspace{0.25cm}\\ \hspace{1.75cm}+ g f^{abc}\overline{\xi}^{a}D^{\m}B^a_{\m}c^{c} -\Omega^{a}_{\m}D^{\m}c^{a} - \Theta^{a\,\s}\left( D_{\s}\xi^{a} + g f^{abc}B^b_{\s}c^{c}\right) \vspace{0.25cm}\\
\hspace{1.75cm} + \dfrac{g}{2}f^{abc}L^{a}c^{b}c^{c} + gf^{abc}\eta^{a}c^{b}\xi^{c} + \dfrac{1}{2}(j+m)\Lp\!\g {A^{a}}_{\m}A^{a\,\m} + {B^{a}}_{\m}B^{a\,\m}\!\Rp+\vspace{0.25cm} \\ \hspace{2.0cm} + \lambda(D_{\m}\xi)^{a}B^{a\,\m} + \gamma\lambda(\partial_{\m}c)^{a}A^{a\,\m} + \tau_{1}m^2j+\dfrac{\tau_{2}}{2} m j^{2}
+ \dfrac{\tau_{3}}{6}  j^{3}\biggr\}
\end{array}
\label{acaototal}
\end{equation}

\noindent Where $\gamma$ and $\tau_{i}$ are parameters.   

\noindent\hspace{0.75cm}Following the procedures presented in section II, we must collect all symmetries compatible with the quantum action principle. Notice that the gauge fixing choice for the $B^{a}_{\s}$ field is compatible with the insertion equation necessary to control the ${B^{a}}_{\m}B^{a\,\m}$ term. Now, as an important data, follows the quantum numbers of all fields and sources presented in the action.

$$
\begin{tabular}{|l|l|l|l|l|l|l|l|l|l|l|l|l|}
\hline
\/ & $A_\mu$ & $B_\mu$&  $c$ &  $\xi$ & $\,\bar c$ & $\,\bar \xi$  & $b$ & $h$ & $\,\,\lambda$ & $j$ & $\Omega_\mu$ & $\Theta_\mu$ \\
\hline
Gh. number & \thinspace 0 & \thinspace 0 & \thinspace 1 & \thinspace 1 & $-1$ & $-1$ & 0 & 0 & $-1$
& 0 & $-1$ & $-1$ \\
\hline
Dimension & \thinspace 1 & \thinspace 1 & $\,\,$0 & $\,\,$0 & $\,$1 & $\,$1 & $\,$1 & $\,$1 & $\,\,$1 & 1 & $\,2$ &
$\,2$ \\
\hline
\end{tabular}\\
$$
\begin{center}
{Table2: Ghost number and canonical dimension of the fields and sources.}
\end{center}

\noindent\hspace{0.75cm}The full set of Functional equations that fix the action $\S $ is given by, the Slavnov Taylor equation
\begin{equation}\begin{array}{ll}
\SS (\S ) = \int d^{3}x \Lp \dfrac{\d \S}{\d A^{a}_{\m}}\dfrac{\d \S}{\d \O^{a \, \m}} + \dfrac{\d \S}{\d B^{a}_{\m}}\dfrac{\d \S}{\d \Theta^{a \, \m}}
+ \dfrac{\d \S}{\d c^{a}}\dfrac{\d \S}{\d L^{a}} 
+ \dfrac{\d \S}{\d \xi^{a}}\dfrac{\d \S}{\d \eta^{a}}+ \vspace{0.25cm} \\ \hspace{2.5cm} 
+ b^{a}\dfrac{\d \S}{\d \cb^{a}} + h^{a}\dfrac{\d \S}{\d \overline{\xi}^{a}}
+ (j + m)\dfrac{\d \S}{\d \l}  \Rp = 0,
\end{array}
\label{Slavnov}
\end{equation}

\noindent and the equation that defines the insertion,

\begin{equation}\begin{array}{ll}
\UU(\S ) = \int d^{3}x\left( \dfrac{\d \S}{\d \lambda} + \xi^{a}\dfrac{\d \S}{\d h^{a}} + g^{2}\gamma c^{a}\dfrac{\d \S}{\d b^{a}} \right) = 0. \vspace{0.25cm} \\ 
\end{array}
\label{insert-B}
\end{equation}

\noindent This equation is deeply related to the gauge fixing choice and for the insertions $A^{a}_{\m}A^{a\,\m}$ and $B^{a}_{\m}B^{a\,\m}$ the gauge fixing compatible with the insertion equation \equ{insert-B} is the Landau one for the $A^{a}_{\m}$ field  

\begin{equation}    
\dfrac{\delta \S}{\delta b^{a}} = \partial^{\s}A^a_{\s},  \hspace{1cm} 
\dfrac{\delta \S}{\delta \overline{c}^{a}} 
+ \partial^{\s}\dfrac{\delta \S}{\delta \Omega^{a\,\s}} = 0
\label{GF}
\end{equation}

\noindent and a nonlinear one for the $B^{a}_{\m}$ field. The nonlinear gauge for the $B^{a}_{\m}$ field generate equations for the gauge fixing and anti-ghost that can not be extended to quantum level. The anticommutation relation of the Slavnov-Taylor and the insertion equation generates the two $\delta $ equations 

\begin{equation}
D(\S )= \int d^{3}x \left(c^{a}\dfrac{\d \S}{\d \overline{c}^{a}} 
+ \dfrac{\d \S}{\d L^{a}}\dfrac{\d \S}{\d b^{a}} \right) = 0, 
\hspace{0.5cm}
K(\S )= \int d^{3}x \left(\xi^{a}\dfrac{\d \S}{\d \overline{\xi}^{a}} 
+ \dfrac{\d \S}{\d \eta^{a}}\dfrac{\d \S}{\d h^{a}} \right) = 0. 
\label{SL2R}
\end{equation}

\noindent\hspace{0.75cm}The action has also the property of the nonrenormalizability of the ghost $c^{a}$. This property is represented in the functional way as the ghost equation.

\begin{equation}\begin{array}{ll}
\GG^{a}(\S ) = \Delta^{a};\vspace{0.25cm} \\ 
\GG^{a} = \int d^{3}x \left[ \dfrac{\d }{\d c^{a}} 
+ g f^{abc} \left( \cb^{b}\dfrac{\d }{\d b^{c}} + \overline{\xi}^{b}\dfrac{\d }{\d h^{c}} \right) \right] \vspace{0.25cm} \\
\Delta^{a} = \int d^{3}x \left[g f^{abc}\left(\O^{b \, \m}{A^{c}}_{\m} 
+ \Theta^{a \, \m}{B^{c}}_{\m}  + c^{b}L^{c} + \xi^{b}\eta^{c}\right) \right] .
\end{array}
\label{def-ghost-B}
\end{equation}

\noindent Finally the group structure is represented by the rigid equation:

\eq\ba{ll}
\WW^{a}(\S)  = \int d^{3}\!x\,g f^{abc}\Lp 
A^{b}_{\m}\dfrac{\d \S}{\d\! A^{c}_{\m}} + B^{b}_{\m}\dfrac{\d \S}{\d\! B^{c}_{\m}}+ \O^{b}_{\m}\dfrac{\d \S}{\d \O^{c}_{\m}} + \Theta^{b}_{\m}\dfrac{\d \S}{\d \Theta^{c}_{\m}} \\ \hspace{3.5cm}+ c^{b}\dfrac{\d \S}{\d c^{c}} + \xi^{b}\dfrac{\d \S}{\d \xi^{c}} 
+ L^{b}\dfrac{\d \S}{\d L^{c}} + \eta^{b}\dfrac{\d \S}{\d \eta^{c}} \\
\hspace{3.5cm} + \cb^{b}\dfrac{\d \S}{\d \cb^{c}} 
+ \overline{\xi}^{b}\dfrac{\d \S}{\d \overline{\xi}^{c}} + b^{b}\dfrac{\d \S}{\d b^{c}} + h^{b}\dfrac{\d \S}{\d h^{c}}\Rp = 0, 
\ea\eqn{rigid-group}

\noindent that is responsible for controlling the group structure of the action.

\noindent\hspace{0.75cm}This complete set of equations is not enough to guarantee the stability of the quantum action. The nonlinear gauge fixing choice for the $B^{a}_{\m}$ field makes it inevitable the mix of the source $ \Theta^{a}_{\m} $ with the BRST transformation of the $A^{a}_{\m}$ field. This is due to the fact that is not possible to write a functional equation for the gauge fixing choice for the $B^{a}_{\m}$ and simultaneously an insertion equation to control the insertion. This is a good example of obstruction and the impossibility to implement simultaneously two different symmetries, represented by the insertion equation and the antighost equation for the $\overline{\xi}$. Observing all the requirements for extending the duality procedure at quantum level, we find that is not possible to use the BF+CS with the insertion of $B^{a}_{\m}B^{a \,\m}$ in order to implement duality as some authors have pointed out using different arguments \cite{Rodrigues}. 

\section{Conclusions}

\noindent\hspace{0.75cm}In this letter, we present one method for implementing a duality mechanism that relates the duality itself to the introduction of gauge condensates into fully quantized actions. The implementation of a set of functional equations that controls the condensate and the starting point action are also discussed. The relation of these condensates with the mechanism of duality was understood in the context of the stability of the starting point action under radiative corrections. This method for obtaining duality under gauge actions is self consistent and gives us information about the possibility of extending the duality procedure to any dimension. The case of duality between the NSD and TM models was re-obtained using the method and the limitations in implementing  duality under  BF+CS and topologically massive Yang-Mills  (TMYM) is, discussed.
 
 These examples prove the power of the method in establishing duality for $ 3 $ dimensional gauge theories.  It is also interesting to point out that the possibility of obtaining Yang-Mills type actions from topological ones open to us one possible connection of topological configurations of gauge fields in Yang-Mills actions with solutions of the topological action with insertion terms. These possible relations could have implications that are very interesting  in $ 3 $ and $ 4 $ dimensional actions. The $4$ dimensional duality is under study.

\vspace{.5cm}
\noindent
{\large\bf{Acknowledgements}}
\newline
The Coordena{\c{c}}{\~{a}}o de Aperfei{\c{c}}oamento de Pessoal de N\'{\i}vel
Superior (CAPES) and the SR2-UERJ are acknowledged for the financial support.


\end{document}